\documentclass[conference]{IEEEtran}

\usepackage{cite}
\usepackage{amsmath,amssymb,amsfonts}
\usepackage{booktabs}
\usepackage{graphicx}
\usepackage{float}
\usepackage{multirow}
\usepackage{algorithm}
\usepackage{algorithmic}
\usepackage{url}
\usepackage{balance}
\usepackage{amsthm}
\usepackage{xcolor}
\newtheorem{lemma}{Lemma}
\newtheorem{theorem}{Theorem}
\newtheorem{proposition}{Proposition}
\newtheorem{remark}{Remark}

\begin{document}

\title{SAGE: Optimal-Stopping Peer Selection for Decentralised Federated Learning}

\author{
\IEEEauthorblockN{Ke Xiao, Qiyuan Wang, Christos Anagnostopoulos}
\textit{School of Computing Science, University of Glasgow, Glasgow, UK}\\
\{ke.xiao, qiyuan.wang, christos.anagnostopoulos\}@glasgow.ac.uk}

\maketitle

\begin{abstract}
Decentralised federated learning replaces server aggregation with peer-to-peer model exchange, making collaborator selection a local decision under uncertainty. Fixed probe budgets waste effort on easy choices yet fall short when peers are hard to distinguish. We propose \textbf{SAGE} (Sequential Anchor-Gated Exchange), an optimal-stopping peer selector under a one-model-bearing-exchange budget. A receiver scores candidate neighbours on receiver-owned anchor evidence and selects once an advantage is certified. It continues probing only while further evidence repays its cost, and otherwise falls back to random gossip. We show that the stopping problem admits an optimal rule attained at a finite stage, and that the anchor schedule is order-optimal in the peer-risk gap and the confidence level. We further show that the selector never returns a peer worse than random gossip with high probability, and prove that no such guarantee holds for selectors that commit without a certificate. A separability threshold follows, below which no probing budget improves on gossip. Experiments span two image benchmarks, two graph families and three heterogeneity levels. Selectors that always act on their evidence lose to gossip in every configuration tested. \textsc{SAGE-OS} matches gossip on $75.5\%$ less evidence than a fixed budget, at half the communication overhead of two published selectors. The operative decision is not which peer to rank first, but whether the evidence justifies ranking at all.
\end{abstract}


\section{Introduction}
Federated learning typically delegates communication and aggregation to a server~\cite{mcmahan2017communication}. In decentralised FL (DFL) there is no such coordinator: clients interact over a graph and each must choose its own collaborator from a local neighbourhood~\cite{lian2017decentralized,yuan2024decentralized}. Under non-IID data that choice is consequential, because a client that spends its one communication opportunity on an incompatible neighbour raises its local loss rather than lowering it.

The choice is also sharply asymmetric in cost. Exchanging a model costs the full parameter payload, whereas scoring a candidate on a handful of held-out examples costs only a few predictions. A client can therefore afford to look before it commits --- but only if it knows how long to look.

Existing peer selectors rank collaborators by local loss, collaborator utility, task or model similarity, selection frequency, or communication cost~\cite{onoszko2021pens,sui2022federico,fan2025pfeddst,soltani2024dflstar,masmoudi2025ocdfl}, and implicitly assume that one fixed evaluation budget is the right amount of evidence for every decision. It is not. An easy neighbourhood is settled by a handful of examples. An ambiguous one may never resolve. Evidence spent there is wasted twice: once in probe cost, and again if the client acts on a ranking the evidence does not support. We therefore ask a different question: not \emph{which peer ranks first}, but \emph{whether the available evidence justifies ranking at all}. Below an explicit threshold on neighbourhood separability, no probing budget improves on random gossip. A receiver starts with a small receiver-owned anchor batch and queries candidate neighbours for bounded loss scores or predictions. In the privacy-preferred form only anchor inputs are sent, labels remain at the receiver, and losses are computed locally from the returned predictions. The batch is enlarged until the best candidate is statistically separated from the runner-up; if confidence is not reached within the probe budget, the client abstains and falls back to random gossip. Only the selected peer transmits the model-bearing object.

We make four contributions.

First, we formulate peer selection as sequential evidence acquisition under an explicit outside option. A receiver faces a three-action decision at every stage --- commit to a peer, buy more evidence, or fall back to gossip --- and we cast this as a finite-horizon optimal-stopping problem under a one-model-bearing-exchange constraint. To our knowledge this is the first treatment of decentralised collaborator selection in which the amount of evidence, rather than the ranking rule, is the object of optimisation.

Second, we show that the resulting anchor schedule is order-optimal. Evidence demand scales as $\Theta(M^{2}\Delta_k^{-2}\log(1/\delta))$ in the risk gap separating the best neighbour from its closest competitor. We establish both directions of this bound. No selection rule, however adaptive, can certify the best peer from fewer anchors.

Third, we establish the conditions under which informed selection is worth attempting at all. We prove that a receiver following our rule never returns a peer worse than random gossip with high probability. No comparable guarantee holds for selectors that commit to the empirically best peer without a certificate. This asymmetry governs their relative behaviour in the experiments. We further identify an explicit separability threshold below which \emph{no} probing budget can improve on gossip, and show that denser neighbourhoods should abandon probing earlier than sparse ones.

Fourth, we evaluate the method against random gossip, fixed-budget anchor selection, two published peer selectors, and two ablations of our own design, across twelve configurations spanning two datasets, two graph families, and three heterogeneity levels. Selectors that always act on their evidence are beaten by uniform gossip on every metric in every configuration. Our stopping rule recovers gossip-level performance on roughly a quarter of the evidence a fixed budget spends, at half the communication overhead of the published selectors.

\section{Related Work}
Decentralised optimisation and gossip learning replace the server with local mixing over graph edges~\cite{lian2017decentralized,ormandi2013gossip,hegedus2019gossip}.  Communication-efficient approaches additionally sparsify or compress the exchanged information~\cite{koloskova2019choco}. 
SAGE is orthogonal to these mechanisms: it asks which feasible neighbour should receive the single model-bearing interaction.

Recent DFL methods increasingly make collaboration selective. PENS evaluates peer models on local data~\cite{onoszko2021pens}; FedeRiCo estimates collaborator utility~\cite{sui2022federico}; PFedDST combines task similarity, loss, and selection frequency~\cite{fan2025pfeddst}; DFLStar selects informative neighbours using last-layer similarity~\cite{soltani2024dflstar}; and OCD-FL balances knowledge gain against communication/energy cost~\cite{masmoudi2025ocdfl}. More recent work learns or adapts the collaboration structure itself: DPFL constructs resource-constrained collaboration graphs~\cite{kharrat2025dpfl}, DFedPGP uses directed asymmetric collaboration~\cite{liu2024dfedpgp}, DFedMQ jointly adapts collaborator selection and topology~\cite{jiang2025dfedmq}, and CFNS combines multiple factors for neighbour selection under local differential privacy~\cite{guo2024cfns}. DA-DPFL instead reduces training and communication cost through dynamic sparse aggregation~\cite{long2025dadpfl}. These approaches address \emph{whom/how to collaborate} or \emph{what to exchange}; SAGE addresses the complementary question of \emph{how much pre-exchange evidence to acquire before committing to one peer}, with statistically certified selection, explicit abstention to gossip, and cost-aware stopping.

Selecting the lowest-risk candidate from a common evidence batch is a pure-exploration problem, and our confidence gate is the decentralised analogue of a racing rule: Hoeffding races prune candidates whose bounds separate~\cite{maron1993hoeffding}, and action elimination and best-arm identification formalise the sample complexity of doing so~\cite{evendar2006action,audibert2010bestarm}. The $\Theta(\Delta^{-2}\log(1/\delta))$ scaling we establish matches the classical bounds for that problem~\cite{mannor2004sample,kaufmann2016complexity}. Two differences matter here. All candidates are scored on the same anchor batch rather than allocated samples adaptively, so the relevant regime is uniform rather than adaptive allocation. The receiver also holds an outside option, gossip, that pure-exploration formulations lack. Abstention rather than identification is therefore the operative decision.

Optimal stopping has also appeared in federated systems, but for different decisions. Dogan-Tusha \emph{et al.} use the secretary problem to select participating IoT nodes from local-model accuracy and received-signal strength~\cite{dogantusha2025ost}, while dynamic hierarchical FL has used optimal stopping to choose relay nodes for energy-efficient parameter upload~\cite{li2023relayost}. These works stop over \emph{which node/relay to accept}; SAGE instead stops over \emph{how much receiver-owned statistical evidence to acquire} before a decentralised peer exchange. To the best of our knowledge, SAGE is the first work to formulate sequential evidence acquisition for peer selection in fully decentralised federated learning as an optimal-stopping problem.

SAGE combines cross-evaluation with lightweight knowledge exchange. PENS demonstrates loss-based peer evaluation~\cite{onoszko2021pens}, while distillation shows that predictions or logits can transfer model knowledge without full-model exchange~\cite{bucila2006model,lin2020feddf}. SAGE keeps model-bearing communication to one peer per round while making the evidence budget adaptive.

\begin{figure}[t]
\centering
\includegraphics[width=0.8\columnwidth]{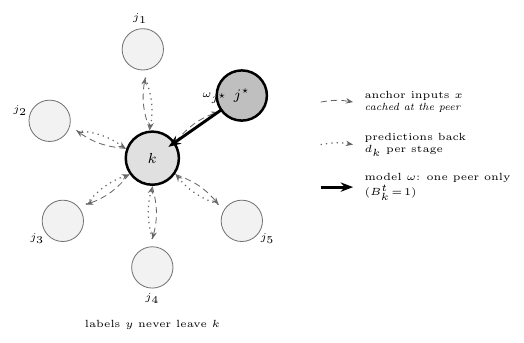}
\caption{Peer selection under a one-model-bearing-exchange budget. Receiver $k$ scores every feasible neighbour on its own held-out anchors, exchanging only inputs and returned predictions, and then admits a full representation from exactly one peer. The probe channel is cheap and reusable; the heavy edge is paid once.}
\label{fig:setting}
\end{figure}

\section{Problem Setting}
Let $\mathcal{G}=(\mathcal{V},\mathcal{E})$ be a communication graph, where $\mathcal{V}=\{1,\ldots,K\}$ is the set of $K$ clients and $\mathcal{E}$ is the set of communication links. Client $k$ has local dataset $D_k$, private data distribution $P_k$, and neighbour set $\mathcal{N}_k$. The round index is $t$, while $j$ indexes a candidate neighbour of receiver $k$. Figure~\ref{fig:setting} illustrates the setting. At round $t$, client $k$ first performs local optimisation and may then initiate a model-bearing peer exchange. Let $B_k^t$ denote the number of neighbours from which client $k$ receives a \emph{model-bearing} object in that round, that is, one carrying model parameters rather than predictions. SAGE studies the communication-constrained setting
\begin{equation}
B_k^t=1,
\label{eq:budget}
\end{equation}
meaning that client $k$ ultimately selects exactly one neighbour for the full representation/model exchange. This is a \emph{design constraint} of SAGE, not a universal assumption of decentralised learning: methods that mix with all neighbours can have $B_k^t=d_k$ (or another value determined by their communication schedule). Importantly, $B_k^t=1$ constrains only the expensive model-bearing exchange. SAGE may still probe multiple candidate neighbours using lightweight evidence before choosing the single peer. Keeping $B_k^t=1$ across the peer-selection baselines isolates whether better selection justifies its probing overhead.

The selected neighbour is $j_k^t\in\mathcal{N}_k$. Let $\Theta_k^t$ denote the complete predictor of client $k$ at the start of round $t$. We write $\Theta_k^{t,+}$ for the model immediately \emph{after local optimisation in round $t$ but before peer mixing}. In the remainder, the superscript ``$+$'' denotes the \emph{post-local-update/pre-mixing} state. After mixing, the resulting shared state is indexed by $t+1$. We decompose the predictor as $\Theta_k=(\omega_k,\phi_k)$, where $\omega_k$ is the shared representation block and $\phi_k$ is a client-local classifier head. Only $\omega$ is model-bearing in the main implementation, while $\phi$ remains local, as in representation-personalised FL~\cite{collins2021exploiting}. After peer selection, receiver $k$ mixes its locally updated representation with that of the selected neighbour:
\begin{equation}
\omega_k^{t+1}=(1-\alpha_{\rm mix})\omega_k^{t,+}+\alpha_{\rm mix}\omega_{j_k^t}^{t,+},
\label{eq:mix}
\end{equation}
where $\alpha_{\rm mix}\in[0,1]$ is the peer-mixing coefficient and $\omega_k^{t+1}$ is the shared block carried into the next round. The peer-selection objective is receiver compatibility rather than necessarily global accuracy. Let $x$ and $y$ denote an input and its label, $\ell(\cdot)$ the probe loss, and
\begin{equation}
R_{k,j}^t=\mathbb{E}_{(x,y)\sim P_k}[\ell(\Theta_j^{t,+};x,y)]
\label{eq:true-risk}
\end{equation}
be the risk of candidate neighbour $j$ on receiver $k$'s distribution after the current local-update phase. Lower $R_{k,j}^t$ means that peer $j$ is more compatible with client $k$ under the chosen probe loss. Client $k$ cannot evaluate~\eqref{eq:true-risk} exactly, so it uses a receiver-owned anchor set $A_k\subset D_k$. For an anchor batch of size $m=|A_k|$, the empirical anchor risk is
\begin{equation}
\widehat R_{k,j}^{t,(m)}=\frac{1}{m}\sum_{(x,y)\in A_k}\ell(\Theta_j^{t,+};x,y).
\label{eq:anchor-risk}
\end{equation}
For readability, when analysing a fixed client-round below we suppress the superscripts $t,+$ and write $R_{k,j}$ and $\widehat R_{k,j}^{(m)}$.

The anchor set $A_k$ is sampled from a held-out local anchor pool disjoint from client $k$'s optimisation and safety-validation data. In the default protocol, client $k$ retains $y$, sends only $x$, and computes the bounded loss locally from each candidate's returned prediction. This avoids disclosing labels or model parameters during probing, although the queried inputs and returned predictions remain exposed. When stronger protection is required, the probing interface can be combined with differential privacy or secure computation techniques, with any induced randomness or approximation incorporated into the confidence analysis. Only the selected peer transmits the model-bearing object.

\section{SAGE: Sequential Anchor-Gated Exchange}
\label{sec:method}
\subsection{Confidence-Gated Selection}
A fixed anchor size, e.g., 32--64 samples, ignores the local decision geometry. `Easy' choices waste probes, whereas ambiguous choices can still be unstable. SAGE instead ties evidence acquisition to ranking confidence. Assume the per-example probe loss is bounded in $[0,M]$, where $M>0$ is a known loss bound. 
A certified choice requires a bounded anchor loss (e.g., classification error); a clipped cross-entropy score can be logged as an additional variant. SAGE uses a nested anchor schedule
\begin{equation}
\mathcal{M}=\{m_1<m_2<\cdots<m_L=m_{\max}\},
\label{eq:schedule}
\end{equation}
where $m_\ell$ is the cumulative number of anchors used at stage $\ell$, $L$ is the number of probing stages, and $m_{\max}$ is the maximum anchor budget per decision.
Because the same decision is inspected at several stages, the confidence bound must account for both candidate peers and repeated looks.

\begin{lemma}[Stage-uniform anchor reliability]
\label{lem:uniform}
Let $d_k=|\mathcal{N}_k|$ be the receiver degree, $L$ the number of anchor stages, and $\delta\in(0,1)$ the per-client-round failure probability. With probability at least $1-\delta$,
\begin{equation}
\left|\widehat R_{k,j}^{(m_\ell)}-R_{k,j}\right|\leq r_\ell
\quad \forall j\in\mathcal{N}_k,\;\ell=1,\ldots,L,
\end{equation}
where the stage-$\ell$ confidence radius is
\begin{equation}
r_\ell=M\sqrt{\frac{\log(2d_kL/\delta)}{2m_\ell}}.
\label{eq:radius}
\end{equation}
\end{lemma}

At stage $\ell$, let $\hat j_1$ and $\hat j_2$ be the lowest- and second-lowest empirical-risk peers. SAGE accepts $\hat j_1$ when
\begin{equation}
\widehat R_{k,\hat j_2}^{(m_\ell)}-\widehat R_{k,\hat j_1}^{(m_\ell)} > 2r_\ell.
\label{eq:stop}
\end{equation}
Otherwise it increases the anchor budget. If no winner is separated by $m_{\max}$, SAGE abstains from deterministic anchor selection and samples a random active neighbour. The confidence parameter $\delta$ is therefore a \emph{per client-round decision} confidence level.

\begin{theorem}[Correct confident selection]
\label{thm:correct}
Whenever condition~\eqref{eq:stop} holds on the event in Lemma~\ref{lem:uniform}, $\hat j_1$ is the true minimum-risk neighbour in $\mathcal{N}_k$.
\end{theorem}

Theorem~\ref{thm:correct} is the main correctness guarantee. The following propositions are finite-sample and stopping results indicating the sufficient conditions and bounds used by SAGE's decision rule. Let $j^\star$ be the true best neighbour and let
\begin{equation}
\Delta_k=\min_{j\neq j^\star}\left(R_{k,j}-R_{k,j^\star}\right)
\label{eq:gap}
\end{equation}
be the true best-versus-runner-up risk gap.

\begin{proposition}
\label{prop:sample}
With probability at least $1-\delta$, SAGE is guaranteed to satisfy the stopping rule once
\begin{equation}
m > \frac{8M^2}{\Delta_k^2}\log\frac{2d_kL}{\delta}.
\label{eq:samplecomplexity}
\end{equation}
\end{proposition}

Equation~\eqref{eq:samplecomplexity} is the main design insight. A decisive neighbourhood ($\Delta_k$ large) needs few probes.  An ambiguous neighbourhood ($\Delta_k$ small) has quadratic probe cost and should often abstain rather than force a ranking. The next result shows that this quadratic dependence is not an artefact of the analysis but intrinsic to the problem: no selection rule, adaptive or otherwise, can certify the best peer with fewer than $\Omega(M^2\Delta_k^{-2}\log(1/\delta))$ anchors.

\begin{theorem}[Anchor lower bound]
\label{thm:lower}
Fix $\Delta\in(0,1/2]$ and $\delta\in(0,1/4)$. Consider a receiver with two candidate neighbours whose per-anchor probe losses take values in $\{0,M\}$, evaluated on a common anchor batch of size $m$. Any selection rule that identifies the lower-risk neighbour with probability at least $1-\delta$ on every instance with risk gap $\Delta$ requires
\begin{equation}
m \;\geq\; \frac{1-\Delta^2/M^2}{8}\cdot\frac{M^2}{\Delta^2}\,\log\frac{1}{4\delta}
\;\geq\; \frac{3}{32}\cdot\frac{M^2}{\Delta^2}\,\log\frac{1}{4\delta}.
\label{eq:lower}
\end{equation}
\end{theorem}

Theorem~\ref{thm:lower} matches Proposition~\ref{prop:sample} in both $\Delta_k^{-2}$ and $\log(1/\delta)$. The residual gap is a constant factor and the additive $\log(d_kL)$ that the union bound of Lemma~\ref{lem:uniform} pays for covering $d_k$ peers and $L$ repeated looks; closing it would require an anytime confidence sequence rather than a stage-uniform one.

\subsection{Optimal Stopping and Gossip Fallback}
The fixed rule ``probe until confidence or $m_{\max}$'' leaves one decision unresolved: an ambiguous client may already know that another probe batch is not worth its cost.  We therefore treat random gossip as an explicit \emph{outside option} and cast anchor acquisition as a finite-horizon optimal-stopping problem, following the standard select/continue logic of optimal stopping~\cite{chow1971great,peskir2006optimal}.
Let $\mathcal{F}_{\ell}$ denote the information available from all probe observations after stage $\ell$. Let $R_{k,\mathrm{RG}}=d_k^{-1}\sum_{j\in\mathcal{N}_k}R_{k,j}$ be the expected anchor risk of choosing uniformly at random from the $d_k$ neighbours, and let $\hat j_{1,\ell}$ be the empirically best peer at stage $\ell$. We express all decision costs in \emph{risk-equivalent units}. If $c_0$ is the raw cost of one anchor--peer evaluation and $\chi>0$ converts one unit of prediction risk into the same raw cost scale, define the single normalized cost parameter
\begin{equation}
\lambda := \frac{c_0}{\chi}.
\label{eq:lambda}
\end{equation}
Thus $\lambda$ is the probing cost of one additional anchor--peer evaluation measured relative to one unit of predictive risk. The conditional costs (superscripts $\mathrm{sel}$, $\mathrm{RG}$, and $\mathrm{cont}$ denote immediate peer selection, immediate random gossip, and continued probing) become
\begin{align}
J^{\rm sel}_{\ell} &= \mathbb{E}[R_{k,\hat j_{1,\ell}}\mid\mathcal{F}_{\ell}],\\
J^{\rm RG}_{\ell} &= \mathbb{E}[R_{k,\mathrm{RG}}\mid\mathcal{F}_{\ell}],\\
J^{\rm cont}_{\ell} &= \lambda d_k(m_{\ell+1}-m_{\ell})+\mathbb{E}[V_{\ell+1}(\mathcal{F}_{\ell+1})\mid\mathcal{F}_{\ell}],
\end{align}
where $V_{\ell}(\mathcal{F}_{\ell})$ is the minimum expected remaining decision cost from stage $\ell$ onward. The optimal value satisfies
\begin{equation}
V_{\ell}(\mathcal{F}_{\ell})=\min\{J^{\rm sel}_{\ell},J^{\rm RG}_{\ell},J^{\rm cont}_{\ell}\},
\label{eq:bellman}
\end{equation}
with $V_L=\min\{J^{\rm sel}_L,J^{\rm RG}_L\}$.  Thus the exact Bayesian optimal policy switches to gossip whenever
\begin{equation}
J^{\rm RG}_{\ell}\leq \min\{J^{\rm sel}_{\ell},J^{\rm cont}_{\ell}\}.
\label{eq:exact-gossip}
\end{equation}
Before using~\eqref{eq:bellman} we record that the problem it describes is well posed: an optimal rule exists, the recursion computes it, and the optimum is attained at a finite stage. The finite horizon makes this elementary: no uniform-integrability or regularity conditions are required beyond boundedness of the probe loss.

\begin{theorem}[Existence of an optimal stopping rule]
\label{thm:exists}
Let the per-example probe loss be bounded in $[0,M]$ with $\lambda,d_k,m_{\max}<\infty$, and let $\Pi$ be the set of policies adapted to $(\mathcal{F}_\ell)_{\ell=1}^{L}$. Then:
\begin{enumerate}
\item[(i)] the recursion~\eqref{eq:bellman} with terminal condition $V_L=\min\{J^{\rm sel}_L,J^{\rm RG}_L\}$ is well defined, and $V_\ell(\mathcal{F}_\ell)$ equals the infimum over $\Pi$ of the expected remaining decision cost from stage $\ell$;
\item[(ii)] the stopping time
\begin{equation}
\tau^{\star}=\min\{\ell\leq L:\ \min\{J^{\rm sel}_\ell,J^{\rm RG}_\ell\}\leq J^{\rm cont}_\ell\}
\label{eq:taustar}
\end{equation}
satisfies $\tau^{\star}\leq L$ almost surely and is optimal, so the infimum in (i) is attained;
\item[(iii)] at $\tau^{\star}$ it is optimal to select $\hat\jmath_{1,\tau^{\star}}$ if $J^{\rm sel}_{\tau^{\star}}\leq J^{\rm RG}_{\tau^{\star}}$ and to gossip otherwise.
\end{enumerate}
\end{theorem}

Theorem~\ref{thm:exists} justifies speaking of \emph{the} optimal policy, but it does not yield an implementable one, because $J^{\rm cont}_\ell$ requires the conditional law of future probe outcomes. Equation~\eqref{eq:bellman} requires a posterior model for future probe outcomes.  SAGE therefore uses a distribution-free sufficient rule derived from the same stopping principle. Let us define the empirical neighbourhood risk and empirical advantage of the currently best peer over random gossip as:
\begin{equation}
\bar R_{\ell}=\frac{1}{d_k}\sum_{j\in\mathcal{N}_k}\widehat R_{k,j}^{(m_{\ell})},
\qquad
A_{\ell}=\bar R_{\ell}-\widehat R_{k,\hat j_{1,\ell}}^{(m_{\ell})}.
\label{eq:gossip-advantage}
\end{equation}
On the event in Lemma~\ref{lem:uniform}, $|R_{k,\mathrm{RG}}-\bar R_{\ell}|\leq r_{\ell}$.

\begin{proposition}[Advantage over gossip]
\label{prop:gossip-cert}
If
\begin{equation}
A_{\ell}>2r_{\ell},
\label{eq:gossip-cert}
\end{equation}
then the empirically best peer is guaranteed to have lower true anchor risk than uniform random gossip:
$R_{k,\hat j_{1,\ell}}<R_{k,\mathrm{RG}}$.
\end{proposition}

Proposition~\ref{prop:gossip-cert} certifies a single decision. Because \textsc{SAGE-OS} selects \emph{only} when that certificate fires and otherwise gossips, the guarantee extends from the certificate to the whole policy.

\begin{theorem}[Gossip safety]
\label{thm:noharm}
On the event of Lemma~\ref{lem:uniform}, the peer $j_k^t$ returned by Algorithm~\ref{alg:sage} satisfies $R_{k,j_k^t}\leq R_{k,\mathrm{RG}}$. Consequently, for a probe loss bounded in $[0,M]$,
\begin{equation}
\mathbb{E}\!\left[R_{k,j_k^t}\right]\;\leq\;R_{k,\mathrm{RG}}+\delta M .
\label{eq:noharm}
\end{equation}
The same bound holds for \textsc{Vanilla SAGE}, whose certificate~\eqref{eq:stop} identifies the true minimiser and therefore also a peer no worse than the neighbourhood mean.
\end{theorem}

\begin{remark}[No such bound exists for always-commit rules]
\label{rem:nocommit}
\textsc{AQ-128} and \textsc{SAGE-NF} return $\arg\min_j\widehat R_{k,j}^{(m)}$ whether or not a certificate fires, so neither branch of the argument applies and the conclusion genuinely fails. When the neighbourhood is near-tied, $\arg\min_j\widehat R^{(m)}_{k,j}$ is determined by sampling noise and lands above $R_{k,\mathrm{RG}}$ with probability approaching that of a uniform draw, a quantity that is bounded by no function of $\delta$. This asymmetry --- a policy-level guarantee for the variants permitted to abstain, and none for those that must commit --- is what separates the two families experimentally.
\end{remark}

The normalized incremental stage cost is
\begin{equation}
C_{\ell+1}^{\rm probe}=\lambda d_k(m_{\ell+1}-m_{\ell}),
\label{eq:stage-cost}
\end{equation}
and the same confidence set upper-bounds the most that perfect peer information could improve upon gossip.
\begin{proposition}[Value-of-perfect-information bound]
\label{prop:vpi}
On the event in Lemma~\ref{lem:uniform},
\begin{equation}
R_{k,\mathrm{RG}}-\min_{j\in\mathcal{N}_k}R_{k,j}
\leq A_{\ell}+2r_{\ell}.
\label{eq:vpi}
\end{equation}
Consequently,
\begin{equation}
C_{\ell+1}^{\rm probe}\geq [A_{\ell}+2r_{\ell}]_{+}
\label{eq:gossip-switch}
\end{equation}
is a sufficient condition for stopping: even perfect information about the best peer cannot repay one more probe stage relative to immediate gossip, where $[z]_{+}=\max\{z,0\}$. If no peer is already certified to beat gossip by~\eqref{eq:gossip-cert}, SAGE-OS therefore switches to random gossip.
\end{proposition}

Two consequences of~\eqref{eq:gossip-switch} characterise \emph{when peer selection is worth attempting at all}. The first is an impossibility statement: below a threshold determined by $\lambda$, the degree, and the first stage of the schedule, no probing budget whatsoever can improve on gossip.

\begin{proposition}[Gossip dominance]
\label{prop:dominance}
Let $G_k=R_{k,\mathrm{RG}}-\min_{j\in\mathcal{N}_k}R_{k,j}$ be the value of perfect peer information. If
\begin{equation}
\lambda\,d_k\,m_1\;\geq\;G_k,
\label{eq:dominance}
\end{equation}
then immediate random gossip is optimal: every policy that acquires any anchor evidence incurs at least as much total cost, regardless of the schedule $\mathcal{M}$, the horizon $L$, or the realised observations.
\end{proposition}

Because $G_k\leq\max_jR_{k,j}-\min_jR_{k,j}$, a receiver can certify dominance from the observed spread alone. Proposition~\ref{prop:dominance} is the theoretical counterpart of the finding in Section~\ref{sec:results} that always-committing selectors are beaten by gossip in every configuration. At the operating point used there ($\lambda=2\times10^{-3}$, $m_1=16$, and $d_k=6$, a representative degree lying between the two graph families) the threshold is $G_k\approx0.19$. Under a bounded $0$--$1$ probe loss, gossip therefore dominates unless mean and minimum neighbour error differ by more than $19$ percentage points. Neighbourhoods that separable are rare, which accounts for the low certified-selection rate and the high gossip rate.

The second consequence formalises the topology effect. The continuation margin is $\mu_\ell(d)=\lambda d(m_{\ell+1}-m_\ell)-[A_\ell+2r_\ell(d)]_+$, with SAGE-OS continuing while $\mu_\ell(d)<0$.

\begin{proposition}[Degree monotonicity]
\label{prop:degree}
Fix a stage $\ell$ and an empirical advantage $A_\ell$. If
\begin{equation}
\lambda\,(m_{\ell+1}-m_\ell)\,d_k\sqrt{2m_\ell\log(2d_kL/\delta)}\;\geq\;M,
\label{eq:degcond}
\end{equation}
then $\mu_\ell$ is strictly increasing in $d_k$, and consequently the stopping stage $\ell^{\star}(d_k)$ is non-increasing in $d_k$: denser neighbourhoods switch to gossip no later than sparser ones.
\end{proposition}

Condition~\eqref{eq:degcond} holds at every stage of the schedule~\eqref{eq:exp-schedule} for $d_k\geq3$, hence throughout both graph families studied here, whose mean degrees are $\approx7.35$ and $\approx5.64$. Proposition~\ref{prop:degree} therefore predicts the ordering observed in Section~\ref{sec:results}, in which the denser Erd\H{o}s--R\'enyi graph triggers cost-based gossip more often than the scale-free graph at every matched dataset and heterogeneity level.

Propositions~\ref{prop:gossip-cert} and~\ref{prop:vpi} therefore act as distribution-free certificates for membership of the stopping region of Theorem~\ref{thm:exists}: whenever~\eqref{eq:gossip-switch} fires, stopping is optimal, so every cost-triggered gossip decision taken by \textsc{SAGE-OS} coincides with an optimal stop. The converse does not hold --- the condition is sufficient, not necessary --- so \textsc{SAGE-OS} may continue at stages where the Bayes-optimal policy would already have stopped. Closing that gap requires exactly the posterior model that~\eqref{eq:bellman} needs and~\eqref{eq:gossip-switch} avoids, so the residual excess is paid in probe cost rather than in selection quality.

Only the ratio $c_0/\chi$ affects the decision, so $\lambda$ removes a degree of freedom while retaining a direct interpretation: larger $\lambda$ makes probing expensive and triggers earlier gossip. Because $C_{\ell+1}^{\rm probe}$ scales linearly with $d_k$, the boundary is topology dependent --- high-degree clients pay more to screen another batch and should gossip earlier unless their peer advantage is correspondingly larger, a prediction the experiments confirm. Condition~\eqref{eq:stop} remains the stronger certificate when the goal is to identify the true best peer rather than merely one better than gossip. Under time-varying availability, $\mathcal{N}_k^t\subseteq\mathcal{N}_k$ is the active feasible neighbour set in round $t$.

\begin{algorithm}[t]
\caption{SAGE-OS at client $k$ in round $t$}
\label{alg:sage}
\begin{algorithmic}[1]
\REQUIRE Active neighbours $\mathcal{N}_k^t$, schedule $m_1<\cdots<m_L$, confidence $\delta$, normalized probe cost $\lambda$
\FOR{$\ell=1,\ldots,L$}
    \STATE Evaluate all active neighbours on the first $m_\ell$ receiver anchors
    \STATE Compute $\widehat R_{k,j}^{(m_\ell)}$, $r_\ell$, $\bar R_\ell$, and $A_\ell$
    \STATE Let $\hat j_1$ be the empirical best peer
    \IF{$A_\ell>2r_\ell$}
        \STATE Select $j_k^t\leftarrow\hat j_1$ \COMMENT{certified better than RG}
        \STATE \textbf{break}
    \ELSIF{$\ell=L$}
        \STATE $j_k^t\sim\mathrm{Uniform}(\mathcal{N}_k^t)$; \textbf{break}
    \ELSE
        \STATE $C_{\ell+1}^{\rm probe}\leftarrow \lambda|\mathcal{N}_k^t|(m_{\ell+1}-m_\ell)$
        \IF{$C_{\ell+1}^{\rm probe}\geq[A_\ell+2r_\ell]_+$}
            \STATE $j_k^t\sim\mathrm{Uniform}(\mathcal{N}_k^t)$ \COMMENT{optimal-stop fallback}
            \STATE \textbf{break}
        \ENDIF
    \ENDIF
\ENDFOR
\STATE Request exactly one model-bearing object from $j_k^t$ and mix using~\eqref{eq:mix}
\end{algorithmic}
\end{algorithm}

\begin{figure}[t]
\centering
\includegraphics[width=\columnwidth]{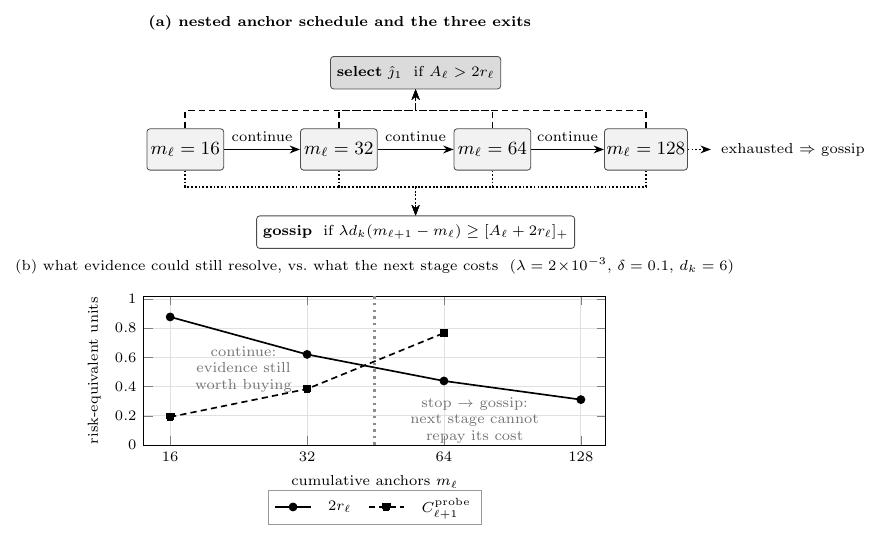}
\caption{The stopping rule. (a) At each stage the receiver may certify a peer, buy the next anchor batch, or abandon probing for gossip. (b) The confidence width $2r_\ell$ bounds what further evidence could still resolve and shrinks as $m_\ell^{-1/2}$, while the cost of the next batch grows with the schedule increment and the degree. Once the curves cross, even perfect peer information cannot repay another stage, which is why terminations concentrate at the middle stages rather than at $m_{\max}$. Panel~(b) is drawn at $d_k=6$, a representative degree between the two graph families used in Section~\ref{sec:experimental-methodology}.}
\label{fig:stopping}
\end{figure}

\paragraph{The proposed method and its ablations.}
\textsc{SAGE-OS}, given in Algorithm~\ref{alg:sage}, is the method we propose. It certifies a peer when that peer is provably better than the random-gossip outside option via~\eqref{eq:gossip-cert}, and switches to gossip \emph{before} $m_{\max}$ whenever the next probing stage cannot justify its cost under~\eqref{eq:gossip-switch}. Two ablations remove one component each, so that the contribution of sequential evidence acquisition, of abstention, and of cost-aware stopping can be separated.

Figure~\ref{fig:stopping} summarises the resulting policy and the quantitative reason it terminates early.

\noindent\textbf{\textsc{Vanilla SAGE}} removes the cost-aware stopping branch. It follows the same nested anchor schedule and the same confidence machinery, selects a peer once the best-versus-runner-up condition~\eqref{eq:stop} is certified, but when confidence is unresolved it \emph{always} advances to the next anchor stage, probing to $m_{\max}$ before abstaining to uniform random gossip. It therefore reaches the same three outcomes as SAGE-OS --- select, or gossip --- but has no way to recognise partway through that a decision will not resolve. \textsc{Vanilla SAGE} versus \textsc{SAGE-OS} isolates the value of cost-aware early stopping.

\noindent\textbf{\textsc{SAGE-NF}} removes abstention instead. It uses the same sequential confidence test as \textsc{Vanilla SAGE}, but if no peer is certified by $m_{\max}$ it deterministically selects the empirical best peer rather than reverting to gossip. \textsc{Vanilla SAGE} versus \textsc{SAGE-NF} isolates the value of declining to act on unresolved evidence.

With the fixed-budget control \textsc{AQ-128}, the four anchor-based selectors form a spectrum. \textsc{AQ-128} always commits on a fixed budget and \textsc{SAGE-NF} always commits on an adaptive one. \textsc{Vanilla SAGE} abstains after exhausting the budget, and \textsc{SAGE-OS} abstains as soon as continuing stops paying.

\section{Experimental Evaluation}
\label{sec:experimental-methodology}
The experiments combine budget-matched controls, which isolate SAGE's adaptive-probing mechanism under $B_k^t=1$, with external peer-selection baselines that test competitiveness with prior DFL methods. Client count, representation model, training horizon, and graph families are shared wherever the comparator permits; any external method with a different communication pattern is measured under that native pattern rather than silently altered.

The evaluation addresses five questions. \textbf{RQ1}: is SAGE-OS competitive with PENS and PFedDST in predictive performance? \textbf{RQ2}: does sequential probing use less evidence than fixed-$m_{\max}$ AQ-128 without degrading performance? \textbf{RQ3}: does cost-aware stopping reduce evidence use beyond confidence gating alone? \textbf{RQ4}: how does graph topology change the value of adaptive selection? \textbf{RQ5}: how do non-IIDness and task difficulty affect evidence demand? Each is settled in the correspondingly tagged subsection of Section~\ref{sec:results}.

\subsection{Experimental Setup}
The study uses Fashion-MNIST (FMNIST) and CIFAR-10 with $K=50$ clients and $T=250$ communication rounds. After the Dirichlet allocation, each client's local data is split into three disjoint parts: $80\%$ for local optimisation, $10\%$ as an \emph{anchor pool} used only for peer probing, and $10\%$ as a local evaluation set used for client-level accuracy reporting. The dataset test set remains shared and is used only for global accuracy and macro-F1. Separating the anchor and local evaluation pools prevents the same samples from both choosing the peer and evaluating client-level performance. Each client uses the same convolutional encoder--decoder representation model as the parent study. FMNIST maps $1\times28\times28$ inputs to a 64-dimensional latent representation with base width 16; CIFAR-10 maps $3\times32\times32$ inputs to a 128-dimensional latent space with base width 32. The classifier head has two fully connected layers. Training uses Adam with learning rate $10^{-3}$, one local epoch per round, and mini-batch size 64. Encoder--decoder weights are model-bearing and are exchanged after peer selection while classifier heads remain local. The peer-mixing coefficient is fixed to $\alpha_{\rm mix}=0.5$, giving equal weight to the receiver and selected peer representation.

Label heterogeneity is generated using a Dirichlet allocation with $\alpha_{\rm Dir}\in\{0.05,0.1,0.3\}$, representing severe, strong, and moderate non-IIDness. These regimes emphasise the setting in which neighbour quality is heterogeneous and informed peer selection can be beneficial. The same client partition is reused across all methods. 
We use two graph families: a connected Erd\H{o}s--R\'{e}nyi (ER) graph with $p=0.15$ (expected degree $\approx7.35$) and a Barab\'{a}si--Albert scale-free graph with $m_{\rm BA}=3$ (realised mean degree $\approx5.64$). Together they contrast relatively homogeneous connectivity with heterogeneous node degree while retaining non-trivial candidate pools for peer selection. 
One graph instance is generated per configuration and reused by every method, and all methods within a configuration share the same client partition and initialisation.
Each reported cell is a single matched run; dispersion across independent seeds is not quantified, and Section~\ref{sec:limitations} states which claims this affects.

\subsection{Baselines}
The anchor schedule, shared by all three SAGE variants, is
\begin{equation}
\mathcal{M}=\{16,32,64,128\},\qquad m_{\max}=128,
\label{eq:exp-schedule}
\end{equation}
with per-decision confidence $\delta=0.10$ and, for SAGE-OS, normalized probing cost $\lambda=2\times10^{-3}$. The anchor subsets are nested, so stage 32 reuses the first 16 anchors and adds only 16 new examples, while later stages reuse all earlier evidence. If a client's anchor pool contains fewer than $m_{\max}$ samples, the schedule is truncated to the largest feasible stage; if fewer than 16 anchors are available, the client falls back directly to RG. Because each client holds only $10\%$ of its local shard as an anchor pool, this truncation is active for a large fraction of clients and is the reason the fixed-budget control does not consume its full nominal budget in Section~\ref{sec:results}. All SAGE variants use the 0--1 classification error as a bounded probe loss for the confidence rule, so $M=1$ exactly.

The \emph{budget-matched core} isolates SAGE under the same one-model-bearing-exchange constraint: \textsc{Local} uses no peer exchange; \textsc{RG} chooses one neighbour uniformly without selection evidence; and \textsc{AQ-128} spends the largest feasible anchor budget on every decision and always selects the empirically best peer. The three SAGE variants are as defined in Section~\ref{sec:method}. The contrast that carries the main claim is \textsc{AQ-128}/\textsc{SAGE-NF} (always commit) against \textsc{Vanilla SAGE}/\textsc{SAGE-OS} (permitted to abstain), with \textsc{RG} as the outside option all four are measured against.

\textsc{PENS}~\cite{onoszko2021pens} is the closest performance-based neighbour selector, using cross-evaluated loss to identify compatible peers. \textsc{PFedDST}~\cite{fan2025pfeddst} combines loss, task similarity, and selection frequency. PENS uses a model-based neighbour-discovery phase and merges multiple peers, so both are measured under their native communication pattern rather than being forced to $B_k^t=1$; their realised exchange cost is reported in Section~\ref{sec:results}.

\subsection{Metrics and Protocol}
We report three metric groups covering learning quality, evidence use, and communication cost.

\textbf{Learning.} Global accuracy and macro-F1 on the shared test set, together with mean client accuracy (Client) and worst-client accuracy (Weak). Both are averaged over clients and are distinct from the \textsc{Local} baseline, which performs no peer exchange.

\textbf{Evidence efficiency.} With final anchor stage $m_{k,t}$ and degree $d_{k,t}$, the saving over a fixed $m_{\max}$ budget is
\begin{equation}
\mathrm{Saving}=1-\frac{\sum_{k,t}d_{k,t}m_{k,t}}{m_{\max}\sum_{k,t}d_{k,t}}.
\label{eq:probesaving}
\end{equation}
Because~\eqref{eq:probesaving} is normalized by the \emph{nominal} $m_{\max}\sum_{k,t}d_{k,t}$, the fixed-budget control \textsc{AQ-128} itself records a non-zero saving that reflects anchor-pool truncation rather than adaptivity; the adaptive gain of SAGE and SAGE-OS is the increment above that value. We report the complementary quantity, \emph{budget consumed} $=100\%-\mathrm{Saving}$, which is normalized by the nominal budget and therefore comparable across all configurations. We additionally report the stopping-stage distribution, the RG-fallback rate, and, for \textsc{SAGE-OS}, the cost-triggered early-gossip rate.

\textbf{Communication.} CommOH is the number of model-bearing exchanges per round relative to \textsc{RG}. All budget-matched methods satisfy $B_k^t=1$ and therefore have $\mathrm{CommOH}=1.000$ by construction; probe traffic is reported separately as budget consumed. PENS and PFedDST exceed $1$ because their native mechanisms exchange models with more than one peer per round.

\paragraph{Configurations.}
Both datasets are run at all three heterogeneity levels under both topologies, giving $12$ configurations and $8$ methods per configuration. All of them are reported; none are omitted.


\section{Results and Discussion}
\label{sec:results}

We report all twelve configurations: two datasets (FMNIST, CIFAR-10), two topologies (ER, scale-free), and three heterogeneity levels ($\alpha_{\rm Dir}\in\{0.05,0.1,0.3\}$). Table~\ref{tab:perf} gives global accuracy and macro-F1, Table~\ref{tab:client} client-level accuracy, and Table~\ref{tab:evidence} evidence use and stopping behaviour. Throughout, budget consumed is $100\%-\mathrm{Saving}$ with Saving as in~\eqref{eq:probesaving}, Stop@$m$ the fraction of decisions terminating at stage $m$, RG fb (RG fallback) the total gossip rate, and Cost-RG the subset triggered by the cost rule~\eqref{eq:gossip-switch}; \emph{always commit} denotes \textsc{AQ-128} and \textsc{SAGE-NF}.

\begin{table*}[t]
\caption{Predictive performance: global accuracy (left) and macro-F1 (right). Bold marks the best method per row and metric.}
\label{tab:perf}
\centering
\tiny
\renewcommand{\arraystretch}{0.9}
\setlength{\tabcolsep}{2.2pt}
\resizebox{\ifdim\width>\textwidth \textwidth\else \width\fi}{!}{%
\begin{tabular}{lll|cccccccc|cccccccc}
\toprule
& & & \multicolumn{8}{c|}{Global accuracy $\uparrow$} & \multicolumn{8}{c}{Macro-F1 $\uparrow$}\\
\cmidrule(lr){4-11}\cmidrule(lr){12-19}
Dataset & Topo. & $\alpha$ & LOCAL & RG & PENS & PFed & AQ-128 & NF & Van. & \textbf{OS} & LOCAL & RG & PENS & PFed & AQ-128 & NF & Van. & \textbf{OS}\\
\midrule
FMNIST & SF & 0.05 & 0.2764 & 0.3039 & 0.3090 & \textbf{0.3099} & 0.2865 & 0.2863 & 0.2971 & 0.2889 & 0.1673 & 0.1882 & \textbf{0.1964} & 0.1962 & 0.1724 & 0.1716 & 0.1809 & 0.1747\\
 & SF & 0.10 & 0.3641 & 0.4075 & 0.4031 & \textbf{0.4102} & 0.3864 & 0.3867 & 0.4075 & 0.4079 & 0.2528 & 0.3005 & 0.2979 & \textbf{0.3049} & 0.2796 & 0.2788 & 0.3019 & 0.3005\\
 & SF & 0.30 & 0.5815 & \textbf{0.6429} & 0.6323 & 0.6420 & 0.6252 & 0.6230 & 0.6391 & 0.6425 & 0.5155 & 0.5843 & 0.5734 & \textbf{0.5847} & 0.5646 & 0.5638 & 0.5799 & 0.5841\\
 & ER & 0.05 & 0.2703 & 0.3114 & 0.3108 & \textbf{0.3142} & 0.2955 & 0.2961 & 0.3083 & 0.3131 & 0.1672 & 0.1967 & 0.1975 & \textbf{0.2003} & 0.1805 & 0.1798 & 0.1948 & 0.1979\\
 & ER & 0.10 & 0.3642 & 0.4104 & 0.4024 & 0.4119 & 0.3826 & 0.3869 & 0.4077 & \textbf{0.4121} & 0.2531 & 0.3041 & 0.2966 & \textbf{0.3053} & 0.2740 & 0.2763 & 0.3019 & 0.3047\\
 & ER & 0.30 & 0.5818 & 0.6430 & 0.6346 & 0.6417 & 0.6293 & 0.6321 & \textbf{0.6473} & 0.6439 & 0.5157 & 0.5845 & 0.5757 & 0.5831 & 0.5692 & 0.5716 & \textbf{0.5899} & 0.5863\\
\midrule
CIFAR-10 & SF & 0.05 & 0.1749 & 0.2089 & 0.1939 & 0.2012 & 0.1725 & 0.1793 & \textbf{0.2091} & 0.2072 & 0.0809 & \textbf{0.1156} & 0.1025 & 0.1072 & 0.0813 & 0.0885 & 0.1143 & 0.1140\\
 & SF & 0.10 & 0.2173 & \textbf{0.2732} & 0.2460 & 0.2593 & 0.2302 & 0.2346 & 0.2621 & 0.2643 & 0.1197 & \textbf{0.1754} & 0.1508 & 0.1607 & 0.1351 & 0.1394 & 0.1644 & 0.1662\\
 & SF & 0.30 & 0.3124 & 0.4260 & 0.3927 & 0.4080 & 0.4048 & 0.4049 & 0.4264 & \textbf{0.4319} & 0.2325 & 0.3610 & 0.3247 & 0.3411 & 0.3386 & 0.3385 & 0.3599 & \textbf{0.3651}\\
 & ER & 0.05 & 0.1744 & \textbf{0.2081} & 0.1917 & 0.2006 & 0.1902 & 0.1966 & 0.2056 & 0.1791 & 0.0811 & \textbf{0.1148} & 0.0988 & 0.1064 & 0.0985 & 0.1033 & 0.1132 & 0.0896\\
 & ER & 0.10 & 0.2176 & 0.2732 & 0.2423 & 0.2667 & 0.2159 & 0.2156 & \textbf{0.2734} & \textbf{0.2734} & 0.1199 & \textbf{0.1763} & 0.1472 & 0.1688 & 0.1206 & 0.1189 & 0.1756 & 0.1748\\
 & ER & 0.30 & 0.3143 & 0.4296 & 0.3917 & 0.4101 & 0.3918 & 0.3948 & 0.4320 & \textbf{0.4355} & 0.2327 & 0.3644 & 0.3239 & 0.3404 & 0.3219 & 0.3245 & 0.3665 & \textbf{0.3683}\\
\midrule
\multicolumn{3}{l|}{\emph{Mean}} & 0.3208 & \textbf{0.3782} & 0.3625 & 0.3730 & 0.3509 & 0.3531 & 0.3763 & 0.3750 & 0.2282 & \textbf{0.2888} & 0.2738 & 0.2833 & 0.2614 & 0.2629 & 0.2869 & 0.2855\\
\bottomrule
\end{tabular}%
}
\end{table*}

\subsection{Committing to the empirically best peer is worse than gossiping (RQ2)}
The principal finding in Table~\ref{tab:perf} is negative. The two always-committing variants, \textsc{AQ-128} and \textsc{SAGE-NF}, are beaten by uniform random gossip in all twelve configurations, with mean accuracy $0.3509$ and $0.3531$ against $0.3782$ for RG. The result is not an artefact of the metric: RG also beats both of them $12/12$ on macro-F1 (Table~\ref{tab:perf}) and $12/12$ on mean client accuracy (Table~\ref{tab:client}). \textsc{PENS} shows the same pattern more mildly, falling below RG in $11/12$ configurations. Theorem~\ref{thm:noharm} and Remark~\ref{rem:nocommit} predict this outcome, and the gap widens where neighbourhoods are hardest to separate.

Spending a full anchor budget and then acting on it is therefore not merely wasteful but actively harmful under $B_k^t=1$. Two mechanisms appear to combine. Anchor risk is a noisy proxy for post-mixing utility, so the arg-min over $\widehat R_{k,j}$ is frequently not the arg-max of realised improvement. In addition, every client applies the same deterministic rule, so exchanges concentrate on a few locally attractive peers and erode the representation diversity that gossip supplies. This justifies treating gossip as an explicit outside option in~\eqref{eq:bellman} rather than a degenerate fallback.

\subsection{Abstention recovers gossip-level performance (RQ2)}
The two variants permitted to abstain close that gap almost exactly, as Theorem~\ref{thm:noharm} requires. \textsc{Vanilla SAGE} reaches mean accuracy $0.3763$, within $0.2$ points of RG, and beats both always-commit variants everywhere on accuracy and macro-F1 alike. \textsc{SAGE-OS} reaches $0.3750$ and is ahead of RG in $7$ of $12$ configurations. Since \textsc{Vanilla SAGE} and \textsc{SAGE-NF} share an identical confidence test and differ \emph{only} in what happens when no peer is certified, the $2.3$-point mean accuracy gap between them ($0.3763$ vs $0.3531$) is attributable entirely to the ability to decline. The value of anchor evidence in this regime is thus largely \emph{diagnostic}: it identifies the decisions in which no peer can be trusted, and its payoff is realised by not acting on them.

\subsection{Optimal stopping makes abstention cheap (RQ3)}
Table~\ref{tab:evidence} quantifies the effect of the cost rule. It beats \textsc{AQ-128} and \textsc{SAGE-NF} in $11$ of $12$ configurations on both accuracy and macro-F1, and matches or exceeds \textsc{Vanilla SAGE} in $8$ of $12$, on far less evidence: mean saving $75.50\%$ against $50.48\%$ and $47.67\%$.

Averaged over the twelve configurations, \textsc{SAGE-OS} concedes $0.13$ accuracy points to \textsc{Vanilla SAGE} and saves $25.0$ points of evidence. The exchange rate is therefore about $190$ points of evidence per accuracy point conceded, which quantifies the effect of cost-aware stopping relative to confidence gating alone.

The moderate-heterogeneity setting exhibits the mechanism most sharply. On FMNIST/scale-free at $\alpha_{\rm Dir}=0.3$, \textsc{Vanilla SAGE} and \textsc{SAGE-OS} both end up gossiping on more than $97\%$ of decisions and land within $0.004$ accuracy of one another ($0.6391$ and $0.6425$). \textsc{Vanilla SAGE} reaches that outcome only after probing to a mean of $78.00$ anchors, exhausting the full budget on $33.9\%$ of decisions. \textsc{SAGE-OS} determines at stage $32$ that no certificate is attainable, stops at a mean of $45.19$ anchors, and never reaches $128$. Both end in gossip, at markedly different cost. Across the full matrix \textsc{Vanilla SAGE} exhausts the budget on $24$--$34\%$ of decisions while \textsc{SAGE-OS} does so on at most $2.0\%$.

\begin{table*}[t]
\caption{Client-level accuracy: mean (Client) and worst-client (Weak). \textsc{Vanilla SAGE} leads on Weak (Section~\ref{sec:results}).}
\label{tab:client}
\centering
\tiny
\renewcommand{\arraystretch}{0.9}
\setlength{\tabcolsep}{2.6pt}
\resizebox{\ifdim\width>\textwidth \textwidth\else \width\fi}{!}{%
\begin{tabular}{lll|ccccc|ccccc}
\toprule
& & & \multicolumn{5}{c|}{Client $\uparrow$} & \multicolumn{5}{c}{Weak $\uparrow$}\\
\cmidrule(lr){4-8}\cmidrule(lr){9-13}
Dataset & Topo. & $\alpha_{\rm Dir}$ & RG & AQ-128 & SAGE-NF & Vanilla & SAGE-OS & RG & AQ-128 & SAGE-NF & Vanilla & SAGE-OS\\
\midrule
FMNIST & SF & 0.05 & \textbf{0.9768} & 0.9702 & 0.9710 & 0.9748 & 0.9731 & 0.6667 & \textbf{0.7576} & 0.7273 & \textbf{0.7576} & 0.6970\\
 & SF & 0.10 & 0.9701 & 0.9617 & 0.9622 & \textbf{0.9702} & 0.9680 & 0.8235 & 0.8235 & 0.8163 & \textbf{0.8824} & 0.8391\\
 & SF & 0.30 & 0.9183 & 0.9091 & 0.9123 & \textbf{0.9239} & 0.9235 & 0.7971 & 0.7869 & 0.7327 & \textbf{0.8143} & 0.7869\\
 & ER & 0.05 & 0.9748 & 0.9733 & 0.9739 & \textbf{0.9755} & 0.9701 & 0.6667 & 0.7273 & \textbf{0.8000} & 0.6667 & 0.6667\\
 & ER & 0.10 & 0.9665 & 0.9518 & 0.9567 & 0.9695 & \textbf{0.9736} & 0.8571 & 0.7059 & 0.7647 & 0.8391 & \textbf{0.8736}\\
 & ER & 0.30 & 0.9190 & 0.9173 & 0.9187 & 0.9206 & \textbf{0.9218} & \textbf{0.8116} & 0.7869 & 0.7705 & 0.7869 & 0.7971\\
\midrule
CIFAR-10 & SF & 0.05 & 0.9166 & 0.8896 & 0.8997 & \textbf{0.9187} & 0.9108 & \textbf{0.5000} & 0.4545 & \textbf{0.5000} & \textbf{0.5000} & \textbf{0.5000}\\
 & SF & 0.10 & 0.8747 & 0.8389 & 0.8401 & \textbf{0.8766} & 0.8706 & 0.2500 & \textbf{0.3333} & 0.2500 & \textbf{0.3333} & 0.2500\\
 & SF & 0.30 & \textbf{0.8142} & 0.7908 & 0.8000 & 0.8049 & 0.8062 & 0.6190 & 0.5227 & 0.5682 & \textbf{0.6349} & 0.6327\\
 & ER & 0.05 & \textbf{0.9226} & 0.8987 & 0.9044 & 0.9126 & 0.9006 & \textbf{0.5000} & \textbf{0.5000} & \textbf{0.5000} & \textbf{0.5000} & \textbf{0.5000}\\
 & ER & 0.10 & 0.8668 & 0.8430 & 0.8189 & \textbf{0.8815} & 0.8684 & 0.3333 & 0.3333 & 0.1667 & \textbf{0.5000} & 0.3333\\
 & ER & 0.30 & 0.8097 & 0.7823 & 0.7820 & 0.8116 & \textbf{0.8205} & 0.5714 & 0.5918 & 0.6122 & \textbf{0.6531} & \textbf{0.6531}\\
\midrule
\multicolumn{3}{l|}{\emph{Mean}} & 0.9108 & 0.8939 & 0.8950 & \textbf{0.9117} & 0.9089 & 0.6164 & 0.6103 & 0.6007 & \textbf{0.6557} & 0.6275\\
\bottomrule
\end{tabular}%
}
\end{table*}

\begin{figure*}[t]
\centering
\includegraphics[width=0.8\textwidth]{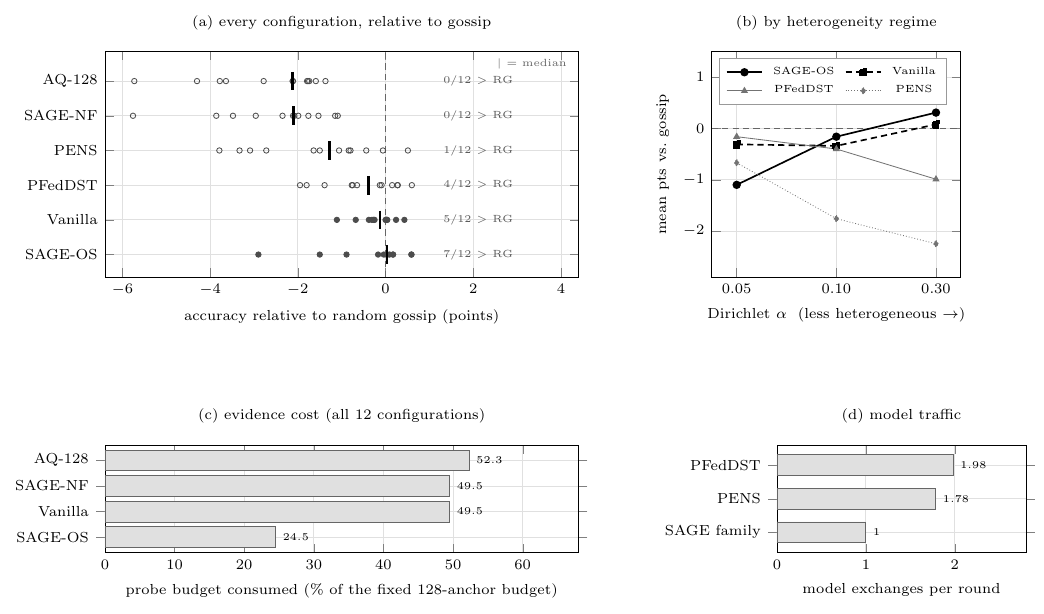}
\caption{Measured outcome and cost, from Tables~\ref{tab:perf}--\ref{tab:evidence}. (a) Accuracy relative to uniform random gossip, one point per configuration, median marked. The two \emph{always commit} selectors are below gossip everywhere; \textsc{SAGE-OS} is above it in seven of twelve. (b) The same quantity by heterogeneity regime: \textsc{SAGE-OS} improves as neighbourhoods become easier to separate, while the published selectors degrade. (c)--(d) Probe budget consumed and model traffic, both normalized counts rather than estimates.}
\label{fig:results}
\end{figure*}

\subsection{Outcome and cost together}
Figure~\ref{fig:results} reports every configuration rather than an average, and four readings follow.

\emph{The split between committing and abstaining is categorical.} The medians of \textsc{AQ-128} and \textsc{SAGE-NF} are $-2.11$ and $-2.09$ points, with worst cases approaching $-5.7$. \textsc{PENS} is above gossip once. \textsc{SAGE-OS} is above gossip in seven of twelve, with a positive median of $+0.03$. Its negative mean comes from a single outlier, examined below, not from a systematic deficit.

\emph{\textsc{SAGE-OS} wins outright more often than any other method.} It is the best of all eight selectors in four of twelve configurations, against three each for gossip, \textsc{PFedDST} and \textsc{Vanilla SAGE}. Macro-F1 agrees: its median is $-0.01$ points, level with gossip, while \textsc{PFedDST} sits at $-0.45$ and the committing variants remain at $0/12$ with medians near $-2.2$.

\emph{The advantage grows as neighbourhoods become separable.} At $\alpha_{\rm Dir}=0.05$ \textsc{SAGE-OS} averages $-1.10$ points against gossip; at $0.10$, $-0.16$; at $0.30$, $+0.31$, winning three of four. The published selectors move the other way, \textsc{PFedDST} from $-0.16$ to $-0.99$ and \textsc{PENS} from $-0.67$ to $-2.25$. Topology shows the same asymmetry: on the denser Erd\H{o}s--R\'enyi family \textsc{SAGE-OS} beats gossip in five of six configurations, against two of six on scale-free graphs. This is the regime Proposition~\ref{prop:degree} concerns, and the direction matches.

\emph{Cost differences are unambiguous.} \textsc{SAGE-OS} consumes $24.5\%$ of the fixed anchor budget, against $49.5\%$ for \textsc{Vanilla SAGE} and \textsc{SAGE-NF} and $52.3\%$ for \textsc{AQ-128}. It issues one model-bearing exchange per round, against $1.784$ for \textsc{PENS} and $1.980$ for \textsc{PFedDST}. Both are normalized counts rather than estimates. Quality differences among the leading methods are smaller than their spread across configurations, so we claim no ordering there. What separates them is cost, and the margin is roughly twofold on both axes.

\subsection{Where Vanilla SAGE remains the better choice}
One axis runs the other way, and it is a genuine trade-off rather than a rounding artefact. On worst-client accuracy (Weak, Table~\ref{tab:client}) \textsc{Vanilla SAGE} is the strongest method in the study, averaging $0.6557$ against $0.6275$ for \textsc{SAGE-OS} and $0.6164$ for RG, and beating \textsc{SAGE-OS} $6$--$4$--$2$ (win--tie--loss); mean client accuracy shows the same weaker ordering ($0.9117$ vs $0.9089$). The clients most in need of a good collaborator are exactly those with the smallest, most skewed shards, for which anchor evidence is noisiest and a certificate emerges only late; stopping at stage $32$ denies them the deeper evidence that would have served them. Macro-F1 shows the same compression, \textsc{SAGE-OS} splitting $6$--$6$ with both \textsc{Vanilla SAGE} and PFedDST against $8/12$ on accuracy.

Where worst-client fairness is the objective and probe budget is not binding, \textsc{Vanilla SAGE} is the appropriate variant. Where the joint performance--cost criterion governs --- the setting this paper targets --- \textsc{SAGE-OS} dominates, trading a $0.03$-point mean accuracy difference for a $25$-point increase in evidence saving.

\begin{table*}[t]
\caption{Probe budget consumed and stopping behaviour. Budget consumed is $100\%-\mathrm{Saving}$ with Saving as in~\eqref{eq:probesaving}; it is normalized by the nominal budget and is therefore comparable across all configurations. Stop@128 is the fraction of decisions exhausting the schedule, RG fb (RG fallback) the total gossip rate, and Cost-RG the subset triggered by the cost rule~\eqref{eq:gossip-switch}.}
\label{tab:evidence}
\centering
\tiny
\renewcommand{\arraystretch}{0.9}
\setlength{\tabcolsep}{3.4pt}
\resizebox{\ifdim\width>\textwidth \textwidth\else \width\fi}{!}{%
\begin{tabular}{lll|ccc|cc|ccc}
\toprule
& & & \multicolumn{3}{c|}{Budget consumed $\downarrow$} & \multicolumn{2}{c|}{Vanilla SAGE} & \multicolumn{3}{c}{SAGE-OS}\\
\cmidrule(lr){4-6}\cmidrule(lr){7-8}\cmidrule(lr){9-11}
Dataset & Topo. & $\alpha_{\rm Dir}$ & AQ-128 & Vanilla & SAGE-OS & Stop@128 & RG fb & Stop@32 & Stop@128 & Cost-RG\\
\midrule
FMNIST & SF & 0.05 & 39.23\% & 35.11\% & \textbf{21.68\%} & 25.6\% & 83.2\% & 26.4\% & 0.0\% & 22.5\%\\
 & SF & 0.10 & 53.77\% & 49.12\% & \textbf{27.71\%} & 30.9\% & 78.4\% & 39.6\% & 2.0\% & 30.5\%\\
 & SF & 0.30 & 62.77\% & 62.59\% & \textbf{29.42\%} & 33.9\% & 97.4\% & 49.6\% & 0.0\% & 52.0\%\\
 & ER & 0.05 & 45.76\% & 42.05\% & \textbf{17.82\%} & 28.0\% & 76.0\% & 50.8\% & 0.0\% & 26.4\%\\
 & ER & 0.10 & 57.84\% & 54.63\% & \textbf{26.14\%} & 32.4\% & 89.3\% & 56.6\% & 2.0\% & 38.8\%\\
 & ER & 0.30 & 62.11\% & 62.04\% & \textbf{27.57\%} & 33.9\% & 97.4\% & 79.7\% & 0.0\% & 71.7\%\\
\midrule
CIFAR-10 & SF & 0.05 & 48.27\% & 43.35\% & \textbf{21.43\%} & 24.2\% & 79.2\% & 36.4\% & 0.0\% & 20.5\%\\
 & SF & 0.10 & 48.49\% & 43.57\% & \textbf{26.15\%} & 24.3\% & 83.9\% & 36.9\% & 0.0\% & 33.5\%\\
 & SF & 0.30 & 55.14\% & 54.98\% & \textbf{27.41\%} & 31.9\% & 95.0\% & 43.0\% & 0.0\% & 44.3\%\\
 & ER & 0.05 & 44.56\% & 38.11\% & \textbf{19.47\%} & 24.2\% & 84.4\% & 46.8\% & 0.0\% & 23.8\%\\
 & ER & 0.10 & 53.92\% & 52.78\% & \textbf{23.38\%} & 31.7\% & 93.6\% & 58.4\% & 0.0\% & 40.1\%\\
 & ER & 0.30 & 56.14\% & 55.88\% & \textbf{25.83\%} & 31.6\% & 95.4\% & 70.1\% & 0.0\% & 56.8\%\\
\midrule
\multicolumn{3}{l|}{\emph{Mean}} & 52.33\% & 49.52\% & \textbf{24.50\%} & 29.4\% & 87.8\% & 49.5\% & 0.3\% & 38.4\%\\
\bottomrule
\end{tabular}%
}
\end{table*}

\subsection{Competitiveness with published selectors at half the exchange cost (RQ1)}
\textsc{SAGE-OS} exceeds \textsc{PENS} in $10/12$ and \textsc{PFedDST} in $8/12$ configurations on global accuracy, and its mean accuracy ($0.3750$) is above both ($0.3625$ and $0.3730$). It does so at $\mathrm{CommOH}=1.000$ against $1.784$ for PENS and $1.980$ for PFedDST, i.e. roughly half the model-bearing traffic. The margin is largest on the harder task: on CIFAR-10/scale-free at $\alpha_{\rm Dir}=0.3$, \textsc{SAGE-OS} reaches $0.4319$ accuracy and $0.3651$ macro-F1 against $0.3927/0.3247$ for PENS and $0.4080/0.3411$ for PFedDST. On FMNIST the three are closer and PFedDST leads in three configurations, \textsc{SAGE-OS} is therefore competitive on the easier task and ahead on the harder one, at half the exchange cost in both.

\subsection{Stopping behaviour follows the theory (RQ4, RQ5)}
Proposition~\ref{prop:sample} predicts that evidence demand scales with the inverse square of the peer-risk gap $\Delta_k$, so certification should become rarer as neighbourhoods become harder to separate. Table~\ref{tab:evidence} confirms this. Proposition~\ref{prop:dominance} set the scale: at this operating point gossip provably dominates unless mean and minimum neighbour error differ by roughly $19$ points, so high gossip rates are the predicted regime rather than a defect. Certified selection --- the complement of the RG-fallback rate --- occurs on roughly $16$--$26\%$ of decisions at $\alpha_{\rm Dir}=0.05$, where sharply skewed label distributions make peers genuinely different, but collapses to $0.2$--$5\%$ at $\alpha_{\rm Dir}=0.3$, where peers are close to interchangeable. \textsc{SAGE-OS} converts that collapse into savings instead of wasted probes: its Cost-RG rate rises monotonically with $\alpha_{\rm Dir}$ in every dataset--topology pair, from $22.5\%$ to $30.5\%$ to $52.0\%$ on FMNIST/scale-free and from $23.8\%$ to $40.1\%$ to $56.8\%$ on CIFAR-10/ER.

Topology acts through degree, also as predicted. Because the stage cost~\eqref{eq:stage-cost} scales linearly with $d_k$, the stopping boundary should bind earlier on denser graphs. It does. At every matched dataset and heterogeneity level, the ER graph (expected degree $\approx7.35$) triggers more cost-based gossip than the scale-free graph (mean degree $\approx5.64$): $38.8\%$ versus $30.5\%$ on FMNIST at $\alpha_{\rm Dir}=0.1$, and $40.1\%$ versus $33.5\%$ on CIFAR-10 at the same level. Savings follow the same ordering, $73.86\%$ versus $72.29\%$ and $76.62\%$ versus $73.85\%$. Neither effect was tuned; both follow from the cost model.

\subsection{Where aggressive stopping hurts}
One configuration departs from this pattern, and we report it as a boundary of the method. On CIFAR-10/ER at $\alpha_{\rm Dir}=0.05$ --- the hardest task at the most severe heterogeneity --- \textsc{SAGE-OS} reaches $0.1791$ accuracy against $0.2056$ for \textsc{Vanilla SAGE}, $0.1902$ for \textsc{AQ-128}, and $0.2081$ for RG, despite recording the largest evidence saving in the matrix ($80.53\%$). This single configuration accounts for essentially the whole difference between the \textsc{Vanilla SAGE} and \textsc{SAGE-OS} column means: excluding it, \textsc{SAGE-OS} averages $0.3928$ against $0.3918$ for \textsc{Vanilla SAGE}.

The diagnosis follows from~\eqref{eq:gossip-switch}. With $\lambda$ fixed, the threshold $C^{\rm probe}_{\ell+1}=\lambda d_k(m_{\ell+1}-m_\ell)$ is scale-free in the probe cost but not in the achievable risk reduction. When the task is hard and the neighbourhood is extremely skewed, the observed advantage $[A_\ell+2r_\ell]_+$ is small at stage $32$, even though a true advantage would have emerged later. The rule is correct given its inputs, but the inputs are uninformative at that stage. A $\lambda$ that adapts to the observed spread of $\widehat R_{k,j}$, or an anytime confidence sequence in place of the stage-uniform bound, is the natural remedy and is left to future work.

\section{Conclusion and Limitations}
\label{sec:limitations}
We introduced SAGE, which casts decentralised peer selection as sequential evidence acquisition, and \textsc{SAGE-OS}, its optimal-stopping instantiation: select once a peer's advantage over gossip is certified, revert to gossip as soon as further probing cannot repay its cost. The stopping problem admits an optimal rule attained at a finite stage, and the anchor schedule is order-optimal in the peer-risk gap and the confidence level. The selector also carries a policy-level guarantee that it never returns a peer worse than gossip. We show that no such guarantee can hold for selectors that commit without a certificate. That asymmetry, with a separability threshold below which no probing budget improves on gossip, predicts what the experiments find. \textsc{Vanilla SAGE} reaches the same decisions by probing to exhaustion, at twice the cost. The operative decision is not which peer to rank first, but whether the evidence justifies ranking at all.

The results also bound the claim: anchor risk is a noisy proxy for post-exchange utility, so \textsc{SAGE-OS} matches gossip cheaply rather than beating it. Each cell is a single matched run, so sub-point margins are ties; only the always-commit-versus-abstain gap ($\approx2.3$ points, $12/12$ in direction on three metrics) and the $\approx25$-point Saving gap survive, and the worst-client lead of \textsc{Vanilla SAGE} is the most seed-sensitive claim reported here. The cost parameter $\lambda$ is fixed and $\delta$, $\mathcal{M}$ unswept; overhead is counted in evaluations rather than seconds;.

\appendices
\section{Proofs}
This appendix collects the proofs of the results stated in Sections~\ref{sec:method}. Numbering follows the main text.

\subsection*{Proof of Lemma~\ref{lem:uniform}}
For a deviation threshold $\epsilon>0$, Hoeffding's inequality gives probability at most $2\exp(-2m_\ell\epsilon^2/M^2)$ for a fixed peer and stage. Setting this to $\delta/(d_kL)$ and applying a union bound over all peers and stages yields~\eqref{eq:radius}.

\subsection*{Proof of Theorem~\ref{thm:correct}}
For any peer $j\neq\hat j_1$, $R_{k,j}\geq \widehat R_{k,j}-r_\ell$ and $R_{k,\hat j_1}\leq\widehat R_{k,\hat j_1}+r_\ell$. Because $\hat j_2$ is the best empirical competitor,~\eqref{eq:stop} implies $\widehat R_{k,j}-r_\ell>\widehat R_{k,\hat j_1}+r_\ell$ for every $j\neq\hat j_1$. Hence $R_{k,j}>R_{k,\hat j_1}$.

\subsection*{Proof of Proposition~\ref{prop:sample}}
On the uniform-confidence event, the empirical gap between the true best peer and any competitor is at least $\Delta_k-2r_\ell$. The stopping rule is therefore guaranteed if $\Delta_k-2r_\ell>2r_\ell$, i.e., $r_\ell<\Delta_k/4$.  Substituting~\eqref{eq:radius} and rearranging yields~\eqref{eq:samplecomplexity}.

\subsection*{Proof of Theorem~\ref{thm:lower}}
Write $\Delta'=\Delta/M$ and construct two instances. Under $P$, neighbour $1$ has per-anchor loss $M\cdot\mathrm{Ber}(\frac{1}{2}-\frac{\Delta'}{2})$ and neighbour $2$ has $M\cdot\mathrm{Ber}(\frac{1}{2}+\frac{\Delta'}{2})$, so neighbour $1$ is the minimiser and the risk gap is $\Delta$; under $Q$ the two are exchanged. Both instances are admissible, so a rule correct with probability $1-\delta$ on each must, writing $A$ for the event that it outputs neighbour $2$, satisfy $P(A)\leq\delta$ and $Q(A^{c})\leq\delta$. The Bretagnolle--Huber inequality gives $P(A)+Q(A^{c})\geq\frac{1}{2}\exp(-\mathrm{KL}(P^{m},Q^{m}))$, whence $\mathrm{KL}(P^{m},Q^{m})\geq\log\frac{1}{4\delta}$. The $m$ anchors are i.i.d.\ and each yields one observation per neighbour, so $\mathrm{KL}(P^{m},Q^{m})=m\left[\mathrm{KL}(\mathrm{Ber}(p),\mathrm{Ber}(q))+\mathrm{KL}(\mathrm{Ber}(q),\mathrm{Ber}(p))\right]$ with $p=\frac{1}{2}-\frac{\Delta'}{2}$, $q=\frac{1}{2}+\frac{\Delta'}{2}$. Using $\mathrm{KL}(\mathrm{Ber}(a),\mathrm{Ber}(b))\leq(a-b)^{2}/(b(1-b))$ and $p(1-p)=q(1-q)=(1-\Delta'^{2})/4$ bounds the bracket by $8\Delta'^{2}/(1-\Delta'^{2})$. Combining, $8m\Delta'^{2}/(1-\Delta'^{2})\geq\log\frac{1}{4\delta}$, which rearranges to~\eqref{eq:lower}; the second inequality uses $\Delta'\leq1/2$.

\subsection*{Proof of Theorem~\ref{thm:exists}}
Every cost in~\eqref{eq:bellman} is bounded in absolute value by $M+\lambda d_km_{\max}$, so all conditional expectations exist and are finite and the recursion is well defined. We induct backwards on $\ell$. At $\ell=L$ continuation is unavailable, so the remaining cost of any policy is $J^{\rm sel}_L$ or $J^{\rm RG}_L$ and $V_L$ is by construction the smaller of the two, establishing (i) at the horizon. Assume (i) holds at stage $\ell+1$. A policy at stage $\ell$ either stops, incurring $J^{\rm sel}_\ell$ or $J^{\rm RG}_\ell$, or continues, incurring the stage cost $\lambda d_k(m_{\ell+1}-m_\ell)$ plus a remaining cost that by the induction hypothesis is at least $\mathbb{E}[V_{\ell+1}\mid\mathcal{F}_\ell]$; in either case its cost is at least $\min\{J^{\rm sel}_\ell,J^{\rm RG}_\ell,J^{\rm cont}_\ell\}=V_\ell$. Conversely the policy that takes a minimising action at stage $\ell$ and behaves optimally thereafter attains $V_\ell$, proving (i) at stage $\ell$. For (ii), the minimising action at $\tau^{\star}$ is by~\eqref{eq:taustar} a stopping action, and $\tau^{\star}\leq L$ because stopping is forced at the horizon; the policy that continues before $\tau^{\star}$ and stops at $\tau^{\star}$ therefore attains $V_1$. Claim (iii) is the identity of the minimiser in~\eqref{eq:bellman} at $\ell=\tau^{\star}$.

\subsection*{Proof of Proposition~\ref{prop:gossip-cert}}
On the uniform-confidence event,
$R_{k,\hat j_{1,\ell}}\leq \widehat R_{k,\hat j_{1,\ell}}+r_{\ell}$ and
$R_{k,\mathrm{RG}}\geq \bar R_{\ell}-r_{\ell}$.  Condition~\eqref{eq:gossip-cert} makes the first quantity strictly smaller than the second.

\subsection*{Proof of Theorem~\ref{thm:noharm}}
Algorithm~\ref{alg:sage} returns a peer in exactly two ways. If it selects at some stage $\ell$, condition~\eqref{eq:gossip-cert} held, and Proposition~\ref{prop:gossip-cert} gives $R_{k,\hat\jmath_{1,\ell}}<R_{k,\mathrm{RG}}$ on the Lemma~\ref{lem:uniform} event. Otherwise it draws uniformly from $\mathcal{N}_k^t$, whose expected anchor risk is $R_{k,\mathrm{RG}}$ by definition. Both branches satisfy the stated inequality. The event holds with probability at least $1-\delta$, and on its complement the risk is at most $M$, giving~\eqref{eq:noharm}. For \textsc{Vanilla SAGE}, Theorem~\ref{thm:correct} gives $R_{k,\hat\jmath_1}=\min_jR_{k,j}\leq R_{k,\mathrm{RG}}$ whenever it selects.

\subsection*{Proof of Proposition~\ref{prop:vpi}}
On the uniform-confidence event,
$R_{k,\mathrm{RG}}\leq \bar R_{\ell}+r_{\ell}$ and
$\min_j R_{k,j}\geq \widehat R_{k,\hat j_{1,\ell}}-r_{\ell}$.  Subtracting yields~\eqref{eq:vpi}.  Any continuation policy must pay at least $C_{\ell+1}^{\rm probe}$ before obtaining additional information, whereas perfect information could reduce the risk relative to gossip by at most the right-hand side of~\eqref{eq:vpi}.  Hence~\eqref{eq:gossip-switch} makes further probing unable to recover its cost.

\subsection*{Proof of Proposition~\ref{prop:dominance}}
Any policy that probes evaluates all $d_k$ active neighbours on at least $m_1$ anchors, incurring cost at least $\lambda d_km_1$ in risk-equivalent units before observing anything. Its expected risk is at least $\min_{j}R_{k,j}$, attained only by an oracle that always selects the true minimiser. Its total cost is therefore at least $\lambda d_km_1+\min_{j}R_{k,j}$, whereas immediate gossip costs $R_{k,\mathrm{RG}}$. Condition~\eqref{eq:dominance} makes the former at least the latter.

\subsection*{Proof of Proposition~\ref{prop:degree}}
Differentiating, $\partial_d[\lambda d(m_{\ell+1}-m_\ell)]=\lambda(m_{\ell+1}-m_\ell)$, a positive constant, while $\partial_d[2r_\ell(d)]=M/(d\sqrt{2m_\ell\log(2dL/\delta)})$, which is positive but $O(1/d)$. Condition~\eqref{eq:degcond} states exactly that the former dominates the latter, so $\mu_\ell$ is strictly increasing in $d_k$. A larger $\mu_\ell$ makes the stopping test fire at the same or an earlier stage, giving monotonicity of $\ell^{\star}$.

\end{document}